# Development and characterization of the efficient portable X-ray imaging device based on Raspberry Pi camera

**Nguyen Duc Ton[a,f], Nguyen Thanh Luan[b,g], Faizan Anjum[a], D. Joseph Daniel[b], Sunghwan Kim[c], Suchart Kothan[d] , Jakrapong Kaewkhao[e], and HongJoo Kim[a,b*]**

[a] *Department of Physics, Kyungpook National University, Daegu, 41566, South Korea*

[b] *Center of High Energy Physics, Kyungpook National University, Daegu, 41566, South Korea*

[c] *Department of Radiological Science, Cheongju University, Cheongju, 28503, Korea*

[d] *Center of Radiation Research and Medical Imaging, Department of Radiologic Technology, Faculty of Associated Medical Sciences, Chiang Mai University, Chiang Mai, 50200, Thailand*

[e] *Center of Excellence in Glass Technology and Materials Science, Nakhon Pathom Rajabhat University, Nakhon Pathom, 73000, Thailand*

[f] *Institute for Nuclear Science and Technology, Vietnam Atomic Energy Institute, Hanoi, 11307, Vietnam*

[g] *Institute of Physics, Vietnam Academy of Science and Technology, Hanoi, 11307, Vietnam*

*E-mail*: hongjoo@knu.ac.kr

ABSTRACT: This study reports the development and characterization of an efficient portable X-ray imaging device built from Raspberry Pi components, including a high-quality 12.3-megapixel camera configured for indirect detection with a $Gd_2O_2S:Tb$ scintillation screen. The device was evaluated under both ambient light and X-ray exposure conditions. Initial characterization under ambient light ensured proper optical focusing; subsequently, camera settings (ISO and exposure time) were evaluated and optimized for X-ray imaging performance. Spatial resolution of the developed device was quantified using the Slanted-Edge method to derive the Modulation Transfer Function (MTF). Beside the low-noise feature, the device achieves $MTF_{20}$ values of 68 lp/mm under ambient light and 25 lp/mm under X-ray irradiation (50 and 70 kV). Moreover, the modularity of the developed device was confirmed by conducting the tests with LYSO:Ce and GAGG:Ce screens. The results demonstrate that this efficient, scientific-grade, compact platform achieves spatial resolution comparable to that of clinical radiography systems, highlighting its potential for applications in scientific, educational, and medical contexts where efficient and portability are critical considerations.

* Corresponding author: hongjoo@knu.ac.kr

## 1 Introduction

X-ray imaging is one of the most transformative techniques in non-destructive testing, providing non-invasive access to the internal structure of objects and biological tissues. Since the first discovery in the 1890s, it has long served as a versatile tool across diverse fields, from industry, security to scientific research, and medicine due to its exceptional penetration capability [1–5]. In industry and security applications, X-ray imaging provides powerful means of non-destructive testing for quality control and safety, allowing internal inspection of complex assemblies such as welds, electronic circuits and composite materials without damaging the object. In airport and cargo security, X-ray scanners help to detect concealed items such as metals, plastics, explosives, and electronics through color-coded images [6–13]. They improve screening efficiency, meet international security standards, and allow real-time inspection of passenger luggage and cargo. For large-scale freight, X-ray imaging systems range from small cabinet scanners to drive-through units for vehicles and containers, incorporating automated detection, material identification, and AI-based image analysis for faster and more reliable screening [10, 12, 14–17]. In healthcare, it enables non-invasive visualization of internal body structures for diagnosing fractures, bone diseases, pulmonary disorders, and foreign objects, while advanced modalities such as computed tomography and mammography further enhance diagnostic precision [5, 18–21]. Besides these practical applications, X-ray imaging also plays a crucial role in scientific research, including crystallography, synchrotron diagnostics, and radiation monitoring. Ongoing advances in detector design, scintillation materials, and computational imaging continue to enhance its resolution, sensitivity, and versatility, broadening its impact across multiple scientific and technological fields [22–33].

Screen film radiography, once dominant, was limited by its narrow dynamic range, chemical processing challenges, and storage difficulties [34–41]. Its replacement by digital radiography brought extended dynamic range, dose efficiency, advanced image processing, and seamless integration with Picture Archiving and Communication Systems [25, 26, 40]. Two main technologies underpin this shift: Computed Radiography (CR), using cassette-based storage-phosphor plates [30, 34, 36, 42–44], and Direct Digital Radiography (DR), employing flat-panel detectors for immediate acquisition [30, 45–52]. Digital radiography employs either indirect conversion, where scintillators such as CsI or $Gd_2O_2S:Tb$ (GOS) convert X-rays into light before photodetection, efficient but leading to light scatter [26, 33, 53, 54], or direct conversion, typically with amorphous selenium (α-Se), which converts X-rays directly to charge for higher spatial resolution [30, 44, 49, 54–56]. Indirect systems dominate general imaging for dose efficiency, while direct systems are preferred for high-resolution applications [24, 54, 57–59]. Despite advances, current systems remain costly, bulky, and inaccessible for small-scale research, education, and resource-limited settings. Achieving low-dose, high-resolution, and large-area coverage in a compact, affordable form remains a challenge [22, 26, 60–62].

Recent progress in compact electronics and open-hardware platforms offers a solution [63–67]. The Raspberry Pi, a low-cost, onboard computer with strong community support, enables modular, portable imaging systems [68–73]. Building on this capability, the device developed in this work was designed for affordability, portability, and ease of use, employing commercially available components to ensure broad accessibility across applications. Its streamlined data-processing workflow, implemented through straightforward algorithms that analyze light intensity from the integrated camera, enables efficient acquisition and interpretation of imaging data, enhancing usability in diverse operational contexts. This work presents a Raspberry Pi-based digital X-ray imaging system in indirect configuration, utilizing a lens-coupled camera, a prism/reflector mirror, and a GOS scintillation screen. The performance evaluation began with a systematic analysis of key imaging metrics, including readout noise, modulation transfer function (MTF), signal-to-noise ratio (SNR), and contrast under varying X-ray conditions, as well as the effects of camera settings such as ISO and exposure time. These findings underscore the feasibility of developing a high-quality, affordable, and portable radiography system for broad use in education, research, and real-world applications.

## 2 Instrument configuration and characterization method

### 2.1 Instrument configuration

The system was designed by following the indirect detection concept, which converts attenuated X-rays into visible light via a GOS scintillation screen. The experimental setup is shown in Fig. 1. A lens-prism optical assembly redirects and focuses the light onto the RPi HQ camera, with a 15×15×15 mm prism deflecting the optical path to protect radiation-sensitive electronics. The indirect imaging process begins when transmitted X-rays, carrying the object's attenuation-dependent intensity pattern, interact with the GOS scintillator screen. The scintillator absorbs the attenuated X-rays and converts their deposited energy into visible light through luminescence, thereby forming a two-dimensional optical image. The emitted light is guided through the optical prism, which deflects it by 90° toward the imaging lens. The “folded” optical path is a key design feature that relocates the radiation-sensitive camera sensor away from the primary X-ray beam, effectively preventing radiation-induced degradation and reducing image noise. Following redirection, an optical lens system collects the redirected photons, focusing them on the Raspberry Pi high-quality camera. The Raspberry Pi computer subsequently controls the camera’s acquisition parameters and captures the resulting digital image for storage and further processing.

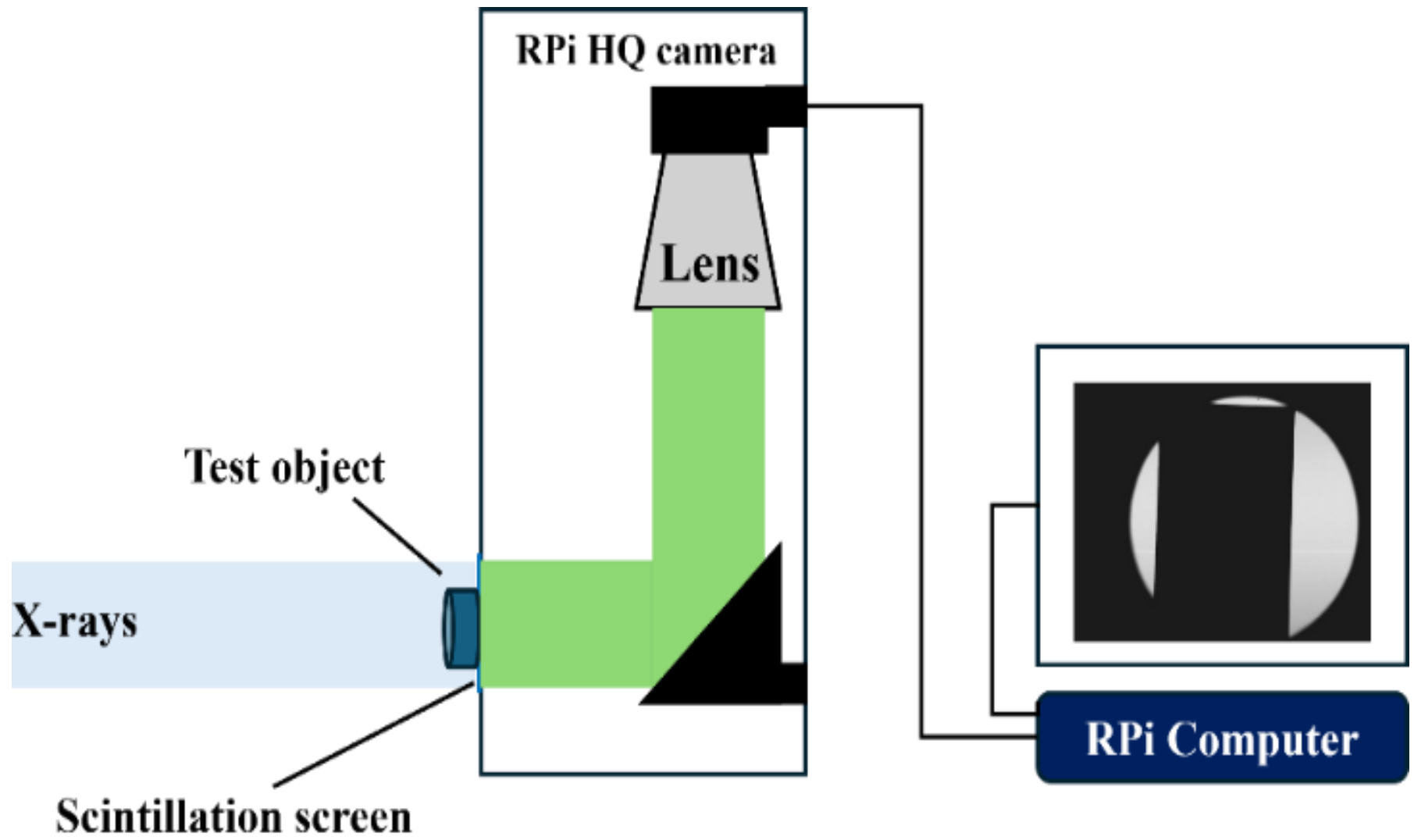

**Figure 1**. Schematic layout of the developed X-ray imaging device in indirect conversion configuration with lens-prism coupled.

The device's central hardware consists of a Raspberry Pi 4 Model B computer paired with its high-quality camera module. While the camera serves as a photosensor, image acquisition and processing are fully controlled by the Raspberry Pi 4B computer, which features onboard storage, multiple connectivity options, and an HDMI output [68, 69]. The imaging sensor is a camera based on a Sony IMX477 sensor, featuring a resolution of 12.3 MP (4056×3040) and a 1.5 μm pixel pitch. With the advantages of CMOS technology, the RPi HQ camera offers relatively low noise, high readout speed, and lower power consumption at a reduced cost. The electronic read noise of RPi HQ camera has been reported to be approximately 3 $e^-$ rms per pixel, placing the Sony IMX477 in the regime of scientific-grade CMOS sensors and enabling low-noise imaging at relatively low signal levels [74]. Moreover, the back-illuminated architecture of the Sony IMX477 sensor relocates the metal interconnects and transistor layers to the rear side of the photodiode, removing these structures from the optical path. This configuration increases the effective fill factor and photon collection efficiency, thereby enhancing quantum efficiency, improving low-light sensitivity, and reducing readout and shot-noise contributions compared with conventional front-illuminated designs [64, 65, 69, 74, 75]. For the indirect imaging setup, it was fitted with a 16 mm CS-mount lens, aligned to an optical prism. Tab. 1 presents detailed information about the RPi HQ camera, which was built on Sony IMX477.

**Table 1**. Basis information of Raspberry Pi High-Quality camera [69, 74, 75].

| Architect | Back-illuminated and staked CMOS image sensor |
|---|---|
| Maximum frame rate | 40 (12 bit-images) |
| Image size | Diagonal 7.857 mm (Type 1/2.3) |
| Number of active pixels | 4056 (H) × 3040 (V) |
| Chip size | 7.564 mm (H) × 5.476 mm (V) |
| Pixel size | 1.55 μm (H) × 1.55 μm (V) |
| Read noise | 3 $e^-$ (rms) |

The camera's interchangeable C/CS mount compatibility and back focusing mechanism allow precise lens-to-sensor adjustment, ensuring sharp focus and adaptability for diverse scientific imaging applications. The sensitivity spectra of the RPi HQ camera based on the Sony IMX477 sensor, as presented in Fig. 2, indicates that the RPi HQ camera achieves peak sensitivity in the 400 - 700 nm range, making it well-suited for indirect detection, where incoming X-rays are converted to visible photons by scintillation screens. Accordingly, a thin GOS screen, characterized by emissions between 450 - 700 nm, was selected for initial evaluations. The X-ray induced luminescence spectrum of GOS was measured and is presented in Fig. 3.

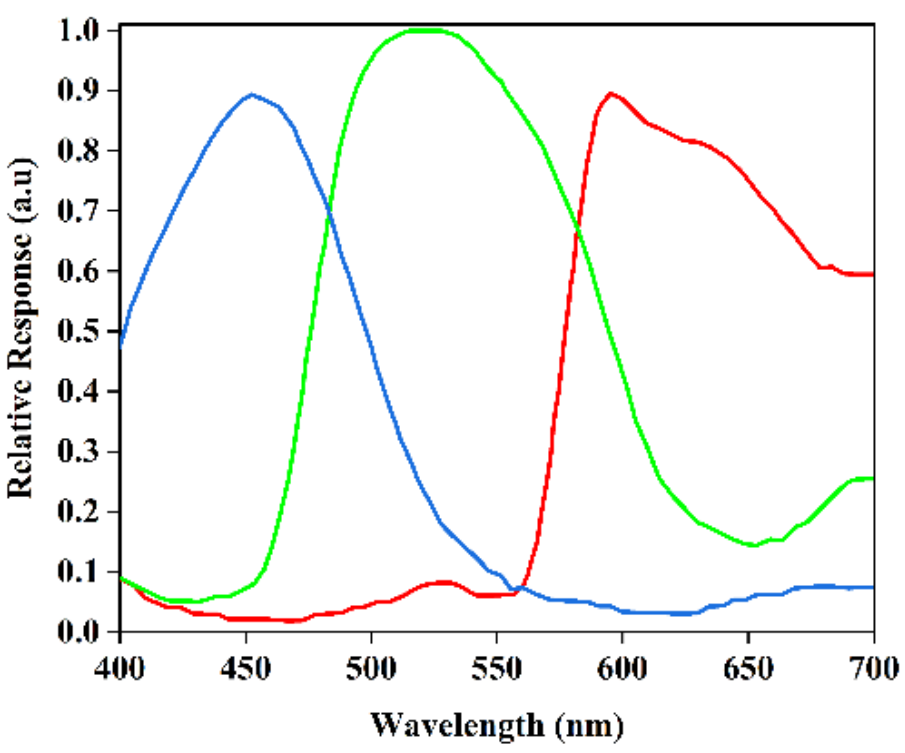

**Figure 2.** Sensitivity spectra of the RPi HQ camera based on Sony IMX477. Data was extracted from [69].

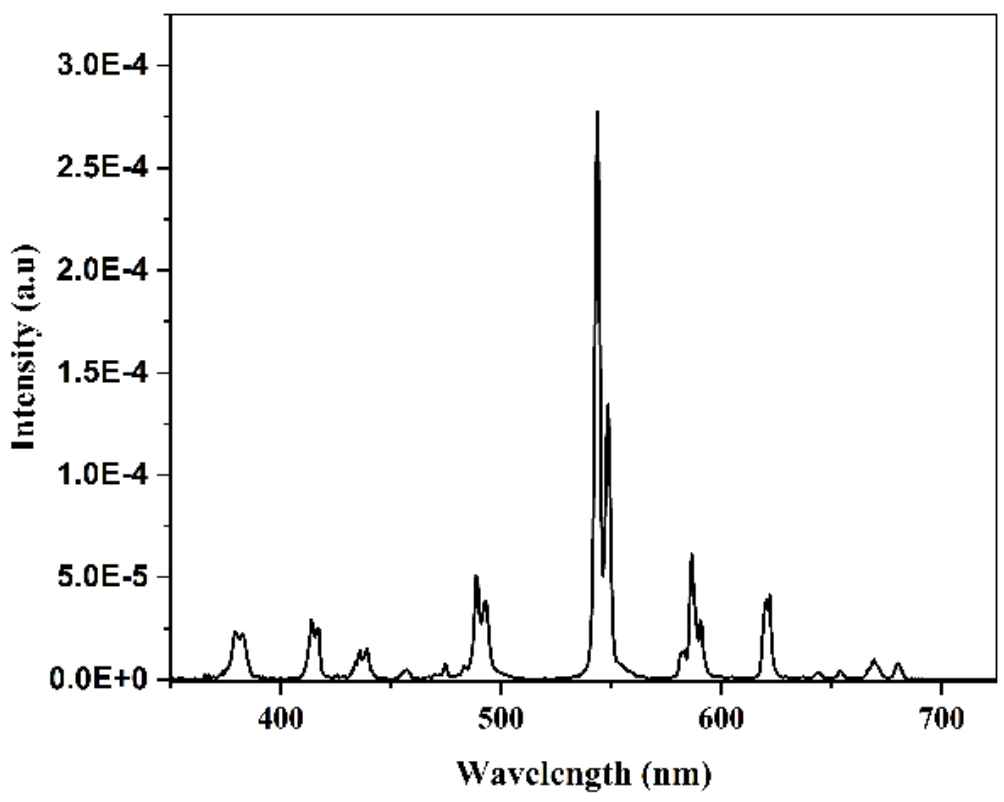

**Figure 3**. Emission spectrum of $Gd_2O_2S$: Tb screen under X-ray excitation.

The Raspberry Pi 4B computer plays the role of the central control and data acquisition unit. The system operates at nominal power consumption of 12–15 W under typical image processing loads, significantly lower than conventional desktop or server-grade systems, enabling integration into compact laboratory setups with minimal thermal management requirements. Remote operation is facilitated through three complementary protocols: (1) SSH for secure command-line access and automated script execution, enabling headless operation and batch processing; (2) VNC for remote graphical desktop access, suitable for real-time image monitoring and parameter adjustment; and (3) Raspberry Pi Connect, a proprietary cloud-based remote desktop service providing intuitive browser-based access without complex configuration. All protocols operate over standard Ethernet or LAN connectivity, accessible from local laboratory networks or external locations via secure connection. From a radiation safety perspective, remote operation implements the ALARA (As Low As Reasonably Achievable) principle's "distance" strategy: by permitting fully control from a separated location, operator proximity to the radiation source is minimized, and cumulative exposure duration is reduced compared to on-site operation. This engineering control measure is particularly valuable for occupational dose minimization in research facilities where imaging system utilization spans extended periods or involves multiple operators, ensuring compliance with regulatory dose limits while maintaining measurement efficiency. For data handling, the system utilizes an SD card for storage and provides multiple connectivity options for straightforward data transmission. A custom Python-based program, built upon standard Raspberry Pi camera libraries and documents, was developed to manage all camera operations and dynamically adjust settings, including exposure mode, ISO, and exposure time [69, 74, 75].

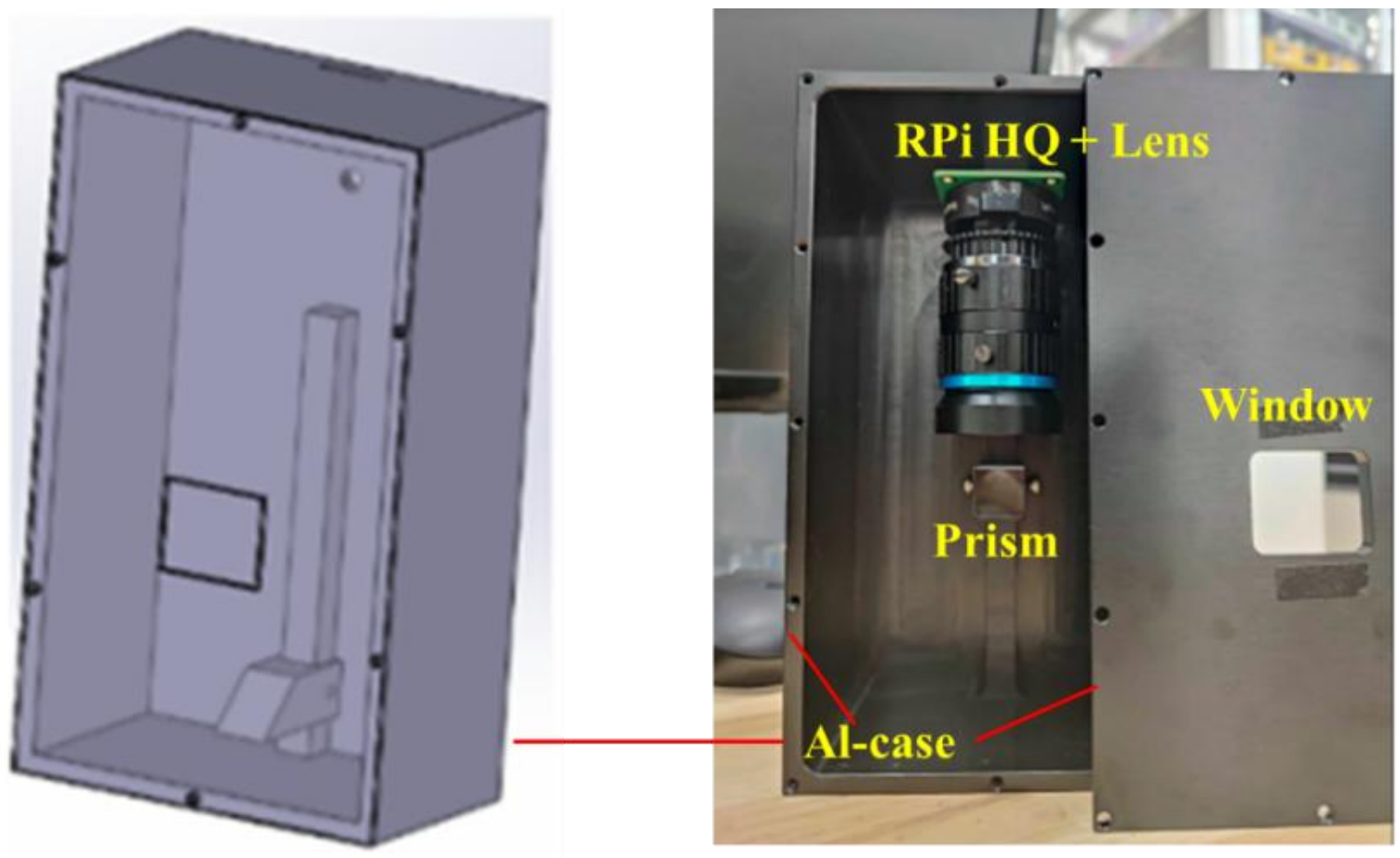

**Figure 4**. (Left) Mechanical drawing of the Aluminum case and (right) the photograph of the optical component installed inside the case.

To achieve the sharp and clear images within a compact imaging system, the optical imaging system was built around a RPi HQ camera and its official 16 mm, 10 MP C-mount lens. The optical path was designed for compactness, featuring a total air-gap distance of approximately 60 mm (from the scintillator to the prism and from the prism to the lens), and a 15 mm thick optical prism. This configuration resulted in an effective object distance that was considerably shorter than the lens's specified Minimum Object Distance (MOD) of 200 mm. A lens could not resolve an image from an object closer than its MOD when set to the standard 17.53 mm C-mount flange distance. To overcome this limitation, the back-focus adjustment ring integrated into the RPi HQ camera was utilized, enabling precise adjustment of the image distance and allowing proper image formation in this compact optical setup. We utilized this mechanism to carefully adjust the lens-to-sensor distance, thereby compensating for the unusually short object distance and achieving a sharp focus on the scintillation screen. For final assembly, the lens's aperture and primary focusing ring were then finely adjusted to achieve the sharpest possible image. Although not inherently designed for low-light applications, the RPi HQ camera achieved effective performance in this setup by operating at a low f-number to enhance light throughput. To reduce ambient noise and shield sensitive components from radiation exposure, the camera sensor, lens, and prism were enclosed in an aluminum housing, as presented in Fig. 4. During experiments, the scintillation screen was attached to the optical window on the cover frame.

To maintain the platform efficient, accessible, and affordable, the system uses readily available components instead of specialized instrumentation or custom-made parts. It builds on the established Raspberry Pi ecosystem with a high-quality camera, together with standard optical and mechanical hardware from high-volume commercial suppliers. This design philosophy promotes wide adoption across educational, research, and resource-limited applications. The total cost of the imaging platform, excluding the X-ray source (typically \$5000-\$50000, available in radiography facilities), depending on type and quality of scintillation screen was estimated approximately \$570. Furthermore, owing to its indirect conversion configuration, the device offers inherent versatility and can be adapted for imaging with other excitation sources, such as neutron or proton beams, by selecting appropriate scintillation materials and optimizing camera parameters.

### 2.2 Characterization method

After fabrication, the system's performance was evaluated at room temperature. The low-noise characteristics of the RPi HQ camera were verified by capturing 10 dark frames at various ISO and exposure settings with the optical window closed, then calculating the standard deviation. The effects of ISO (100 - 800) and exposure time (100 - 2000 ms) on image contrast were assessed at a constant X-ray tube voltage of 70 kV. All images were saved as 12-bit PNG files, in gray scale at the sensor's maximum resolution (4056×3040). Contrast level in percentage was calculated from fixed-size regions of interest (ROIs) in bright and dark regions using:

$$C = \frac{I_{max} - I_{min}}{I_{max} + I_{min}} \times 100\%$$

where $I_{max}$ and $I_{min}$ are pixel intensity in bright and dark areas, respectively. The ROI selection is illustrated in Fig. 5. ROIs were selected as uniform 200×200 pixel areas to minimize artifacts and ensure consistent measurements. This standardized approach was maintained across all measurements for reproducibility. Once the optimal camera settings were determined, the device was then installed for experiments with a standard hospital X-ray head, as shown in Fig. 6, with the source-to-screen distance fixed at 40 cm while the collimators were adjusted to focus X-ray beam to scintillation screen. Images were acquired at various tube voltages (kV), currents (mA), and durations (ms), as listed in Table 2.

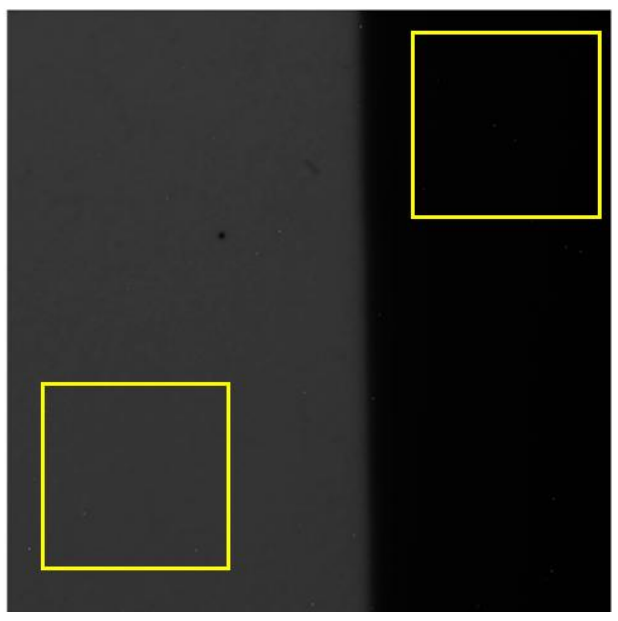

**Figure 5.** An example of ROIs selection for contrast calculation. Two regions were selected on the bright and dark sides to determine the bright and dark levels.

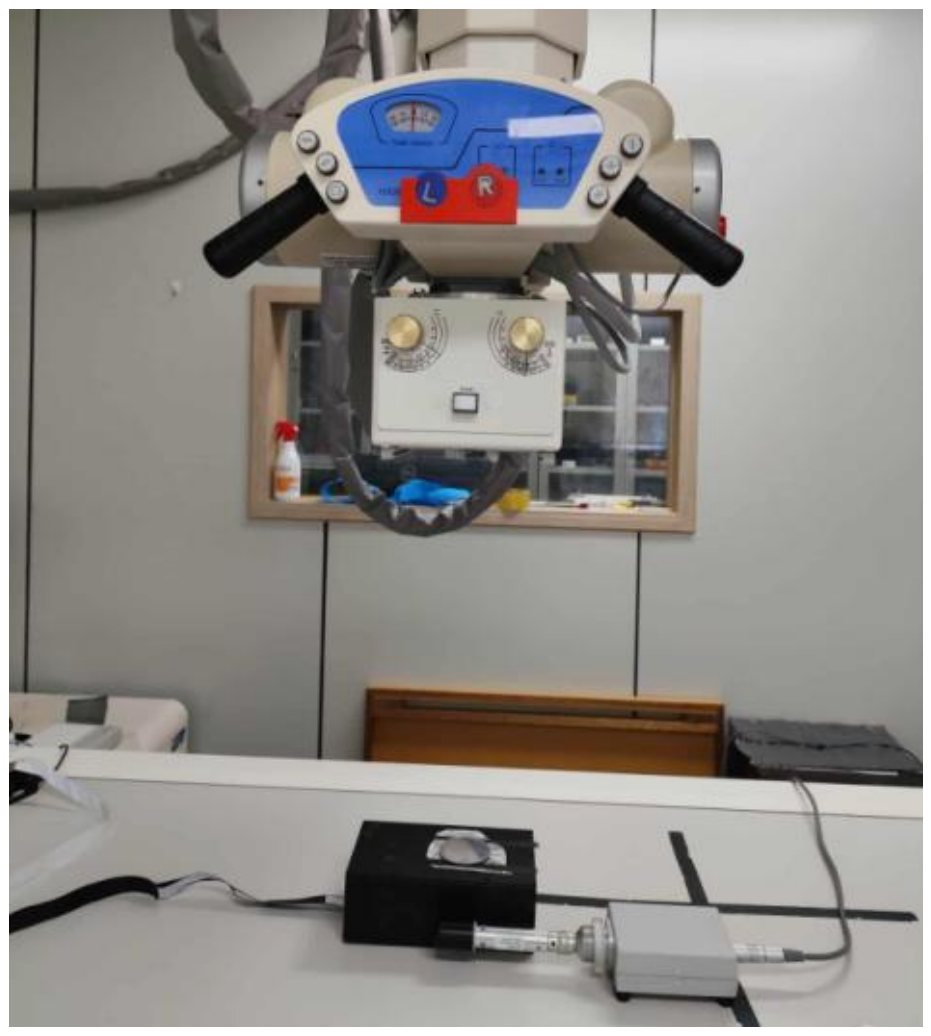

**Figure 6**. Experimental setup with a clinical X-ray source; source-to-screen distance 40 cm. A RadCal Mod 10X5-6 dosimeter measured exposure dose at the scintillator surface.

**Table 2.** Parameters of X-ray beam (kV-mA-ms) at various exposure conditions (kV-mAs) for MTF evaluation.

| mAs/kV | 50 kV | 70 kV |
|---|---|---|
| 5 | 100 mA-50 ms | 100 mA-50 ms |
| 10 | 100 mA-10 ms | 100 mA-10 ms |
| 15 | 100 mA-150 ms | 100 mA-150 ms |
| 20 | 100 mA-200 ms | 100 mA-200 ms |
| 30 | 200 mA-150 ms | 200 mA-150 ms |
| 40 | 200 mA-200 ms | 200 mA-200 ms |

For the preliminary evaluation, the device's performance was assessed using MTF, image contrast, and SNR as a function of exposure dose (mAs or R) at tube voltages of 50 and 70 kV. SNR was computed from a uniform bright ROI by:

$$SNR=\mu/\sigma$$

Where $\mu$ and $\sigma$ are the mean pixel intensity and the standard deviation of the selected ROI, respectively. The consistent size and location across images were optimized to reduce statistical uncertainty and ensure consistency among images. Due to instrumentation limitations, X-ray fluence was not measured; consequently, further performance parameters, including Detective Quantum Efficiency (DQE) and Normalized Noise Power Spectrum (NNPS) as described in IEC 62220-1 standards, would not be accessed [75–79].

Principally, the resolving power of an imaging system at various frequencies, expressed as MTF, could be derived by measuring different spread functions, as illustrated in Fig. 7. The slanted-edge method was employed to evaluate the system's Modulation Transfer Function (MTF). This method is recognized for its high precision, effective noise reduction through averaging, and strong compatibility with digital image processing, making it well-suited for assessing the spatial resolution of digital imaging systems [76–80]. An illustration of the slanted-edge method is shown in Fig. 8. Edge images of the test object were acquired at the camera's maximum resolution and saved into 12-bit PNG files. The method analyzes the system's response to a high-contrast edge positioned at a slight angle to the pixel grid.

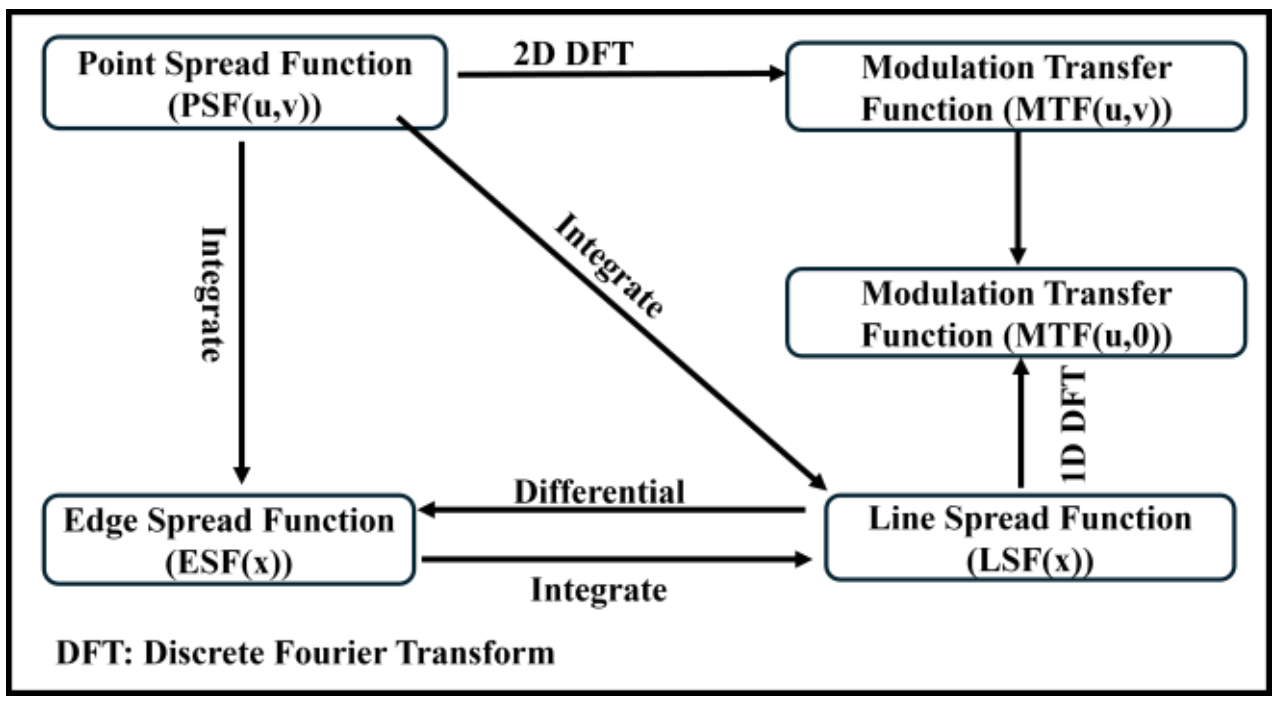

**Figure 7**. Illustration of the relation between the spread functions and the MTF calculations.

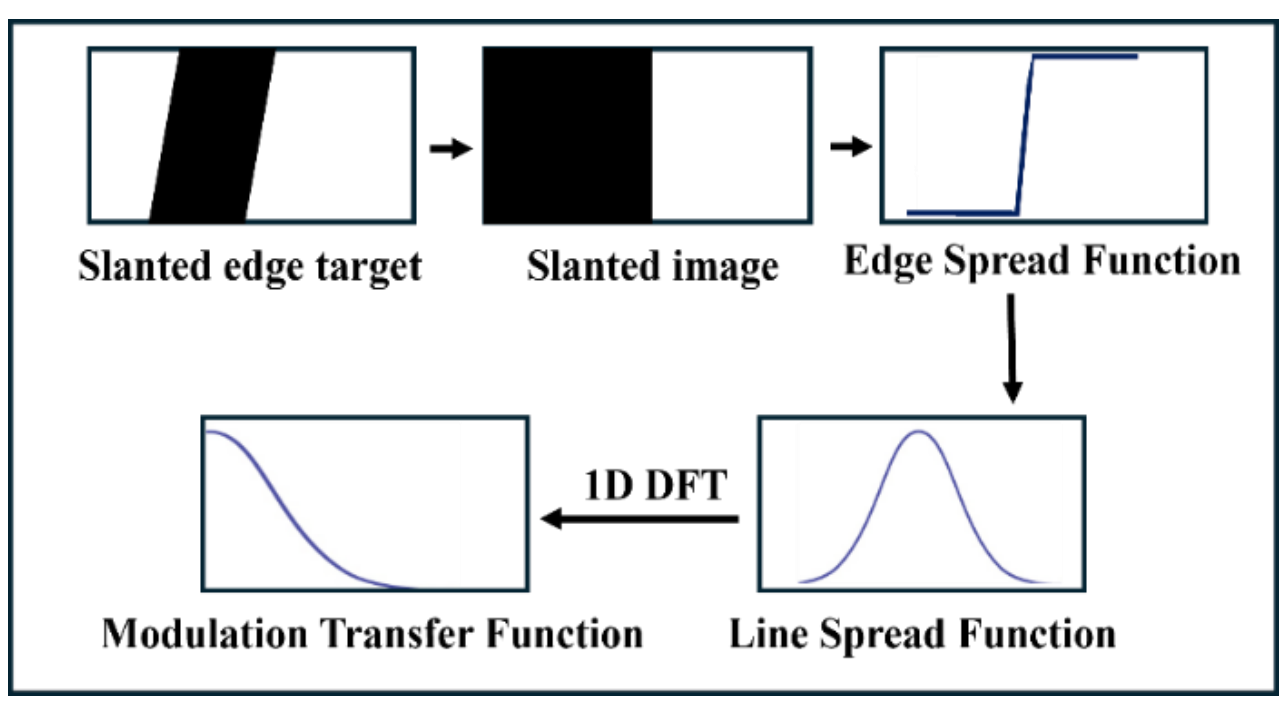

**Figure 8**. Schematic diagram of the Slanted Edge method that was used for MTF measurement.

Following the recommendations from the International Electrotechnical Commission (IEC) to evaluate the performance of X-ray imaging devices, a homemade W plate with 0.1 mm of thickness and a sharp, well-polished edge was attached to the GOS screen at a 2–5° tilted [79, 81, 82]. From the edge images, the Edge Spread Function (ESF) was derived, differentiated to obtain the Line Spread Function (LSF), and Fourier-transformed to yield the MTF [83, 84]:

$$LSF(x) = \frac{d}{dx} ESF(x)$$

$$MTF(u, 0) = |F[LSF(x)]|$$

To validate the MTF extraction, a Teledyne Remote RadEye HR sensor with Min-R ($Gd_2O_2S$:Tb) scintillation film was used [85, 86]. For this setup, the prepared W target was mounted on the sensor's protective layer to acquire X-ray images, from which the MTF curve was extracted and compared with Teledyne's reference data [87].

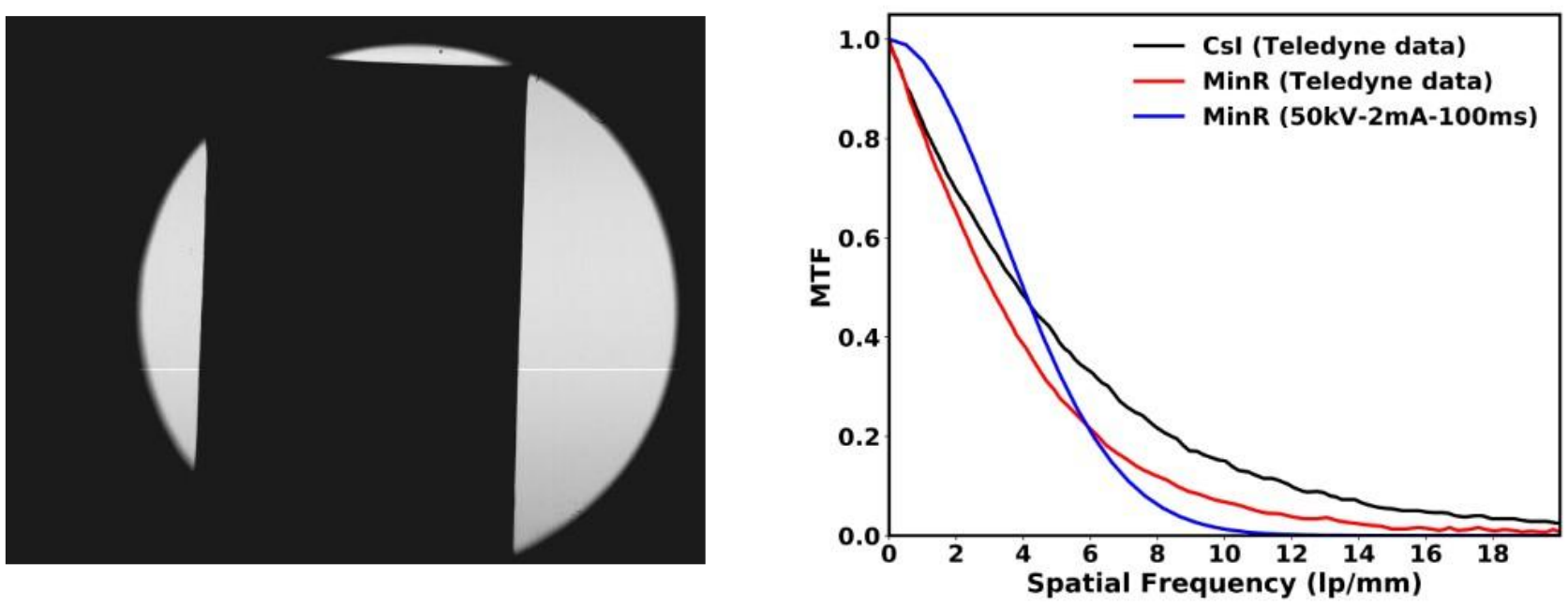

**Figure 9**. (Left) X-ray image recorded by RadEye Remote HR sensor and (right) extracted MTF curve from the captured images (blue line) in compared with its published data (black and red).

The recorded X-ray image and corresponding MTF curve are presented in Fig. 9. The measured MTF exceeds the reference Min-R data at low spatial frequencies, indicating improved performance in that range, but falls off more rapidly beyond the spatial frequency value of approximately 6 lp/mm, reflecting a reduction in high-frequency response. These differences likely stem from variations in experiment conditions and data processing. Overall, the results agree reasonably well with reference data, confirming the reliability of the MTF evaluation. However, the lens-prism coupling inherently reduces the photon throughput, lowering quantum efficiency and SNR as noted in previous studies [21, 88–90]. It should be noted that except read noise measurements, dark-frame correction has been systematically implemented to account for these variations, thereby improving measurement accuracy and uniformity across the sensor during data analysis.

## 3 Results and discussion

### 3.1 Optical performance verification

After assembly, the device's optical alignment and resolution were verified by replacing the scintillation screen with a standard optical resolution target for ambient-light testing. As shown in Fig. 10, bar patterns up to approximately 30-40 lp/mm were clearly resolved, confirming accurate focusing and optical performance. By following the slanted-edge method, the MTF curves under ambient light, which are shown on Fig. 11, yielded approximately 68 lp/mm at 20% of MTF ($MTF_{20}$), demonstrating high spatial resolution and establishing a strong baseline before X-ray imaging.

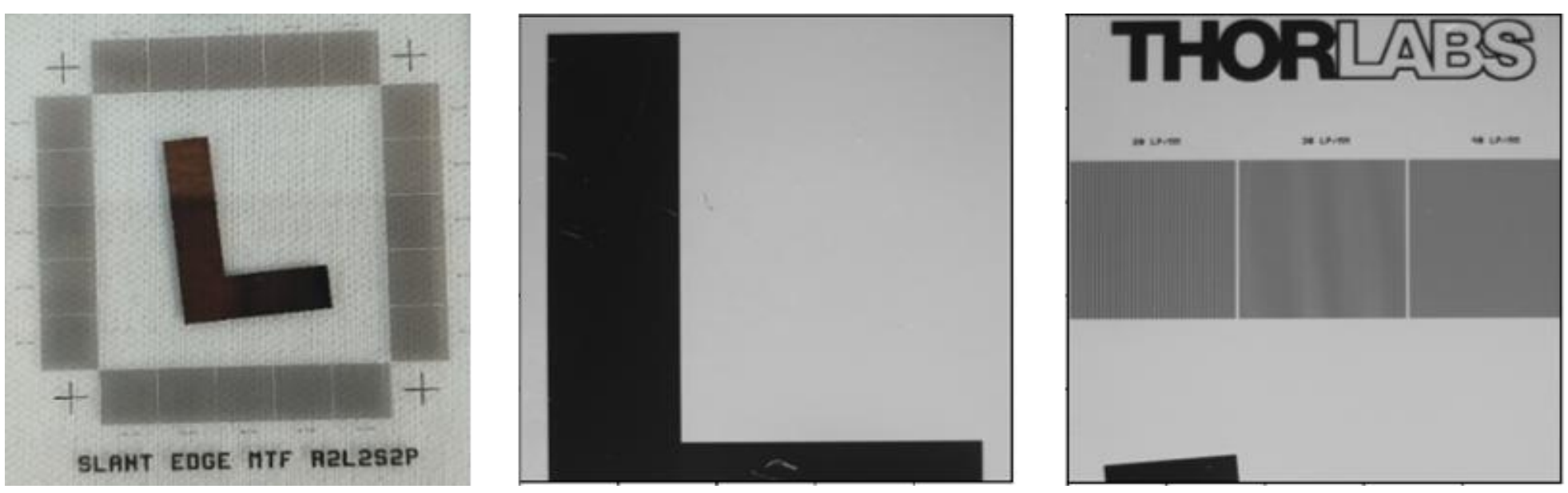

**Figure 10**. The standard resolution target THORLAB R2L2S2P and its optical images were captured by the developed device.

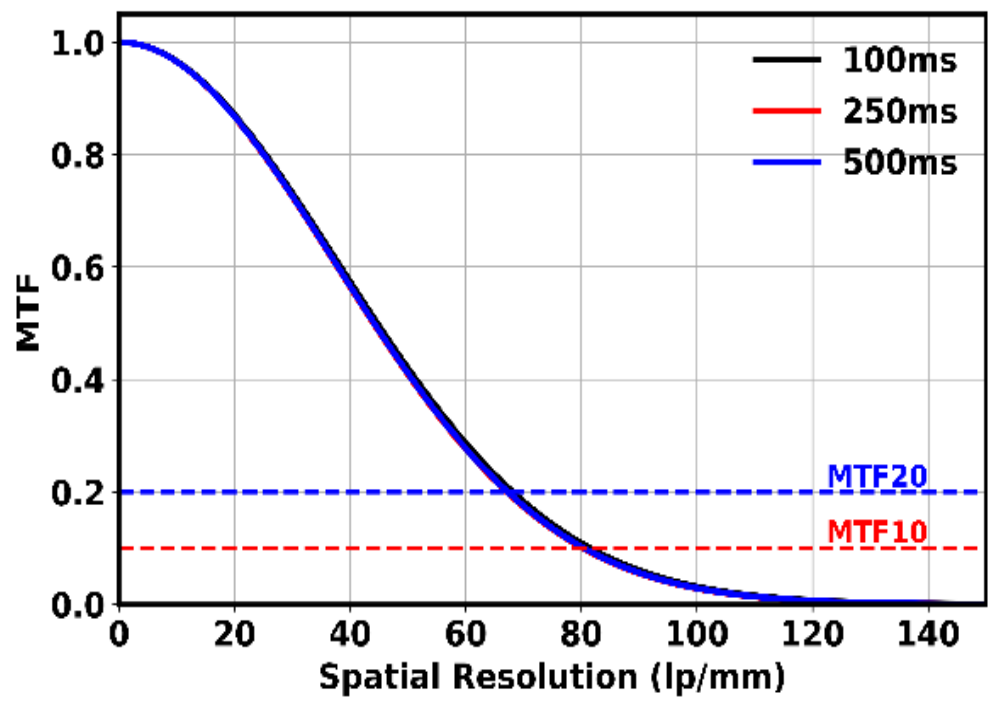

**Figure 11**. MTF curves of the developed device under test with ambient light, ISO100, various exposure times.

### 3.2 Read noise characterization

The read noise or dark noise of the RPi HQ camera was evaluated across ISO (100-800) and exposure times by capturing and averaging five dark frames, then extracting histograms, mean values, and standard deviations.

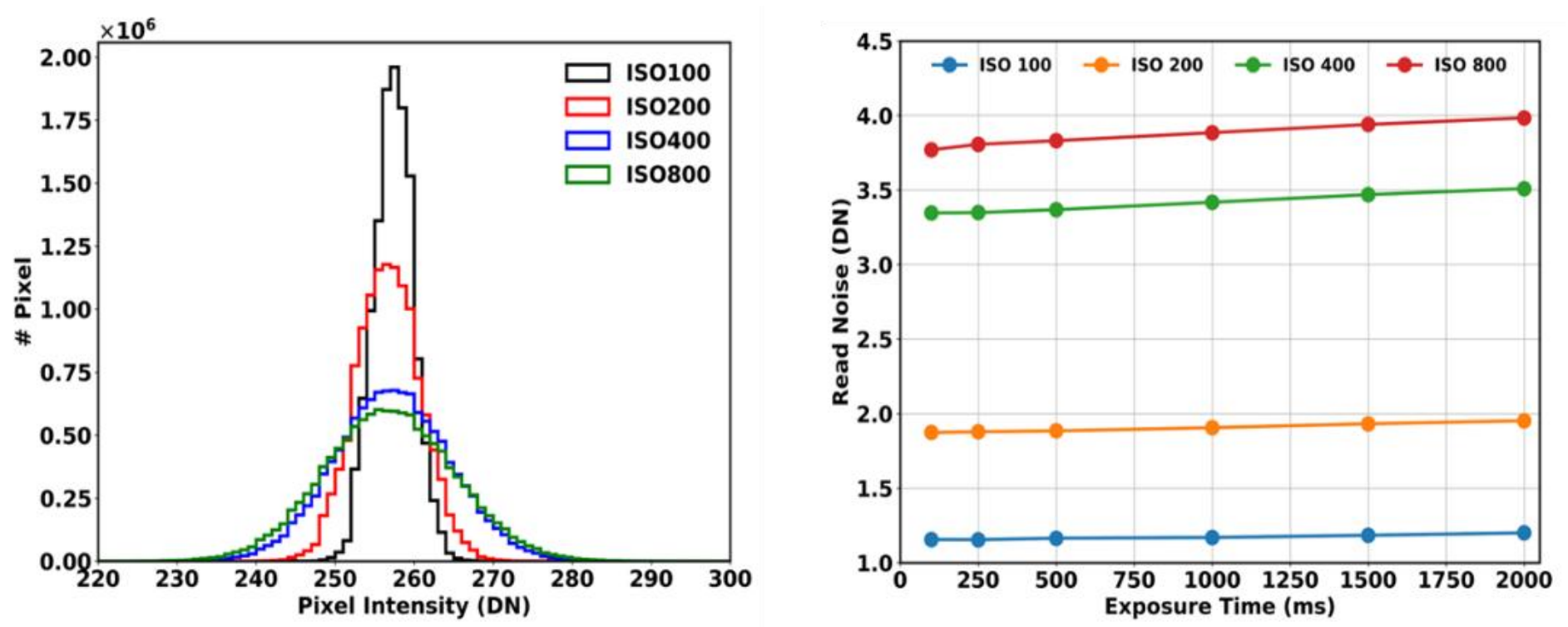

**Figure 12**. (a) Illustration of image histogram of dark image for readout noise evaluation at different ISO settings, 250 ms of exposure time, and (b) measured readout noise (DN) at different ISO settings, exposure times.

As shown in Fig. 12(a), higher ISO broadened and flattened the histograms, indicating increased noise from amplified signal and noise components while the mean values stabilized at 256-257.5. This trend is indicative of increased noise, primarily due to the amplification of both signal and noise components at higher ISO settings. ISO 100 exhibits the narrowest distribution, confirming its superior noise performance. Fig. 12(b) shows that the read noise (standard deviation) increases slightly with longer exposures, particularly at high ISO, suggesting a minor temporal noise component from dark current or thermal effects. ISO gain remained the dominant noise source, with ISO 800 being the highest and ISO 100 being the lowest.

Overall, the RPi HQ camera, which is based on a CMOS sensor, exhibits low intrinsic noise at low ISO settings and stable performance across various exposure times [64, 65, 91]. Even at the highest ISO setting (800) and a long exposure time (2000 ms), the standard deviation is approximately 4, corresponding to less than 1.6 % deviation relative to the mean value of the recorded images. This further demonstrates the low-noise performance of the Raspberry Pi HQ camera. These characteristics are essential for high-fidelity and low-noise imaging, underscoring the suitability of the developed device for precision imaging applications such as X-ray and scientific imaging, where noise performance is critical. Following ambient-light verification, the system was tested in X-ray configuration at 70 kV - 2 mA, with a 3 cm distance between the source and the object. Fig. 13 shows captured images, demonstrating the device's capability for X-ray imaging and its potential for non-destructive testing and inspection. Based on this, further evaluations, optimizations are planned, which will be discussed later.

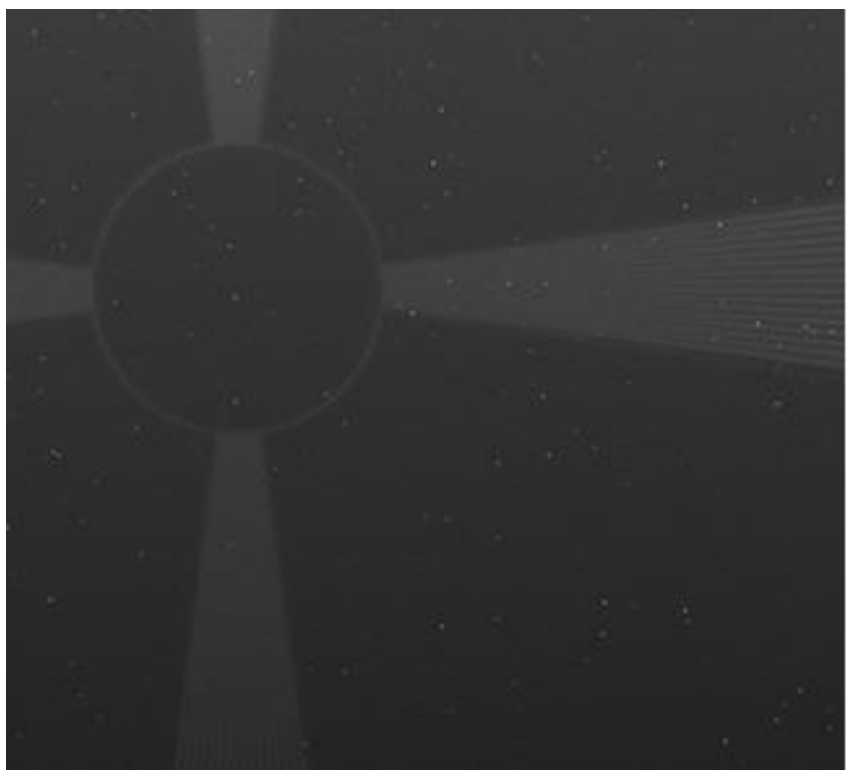

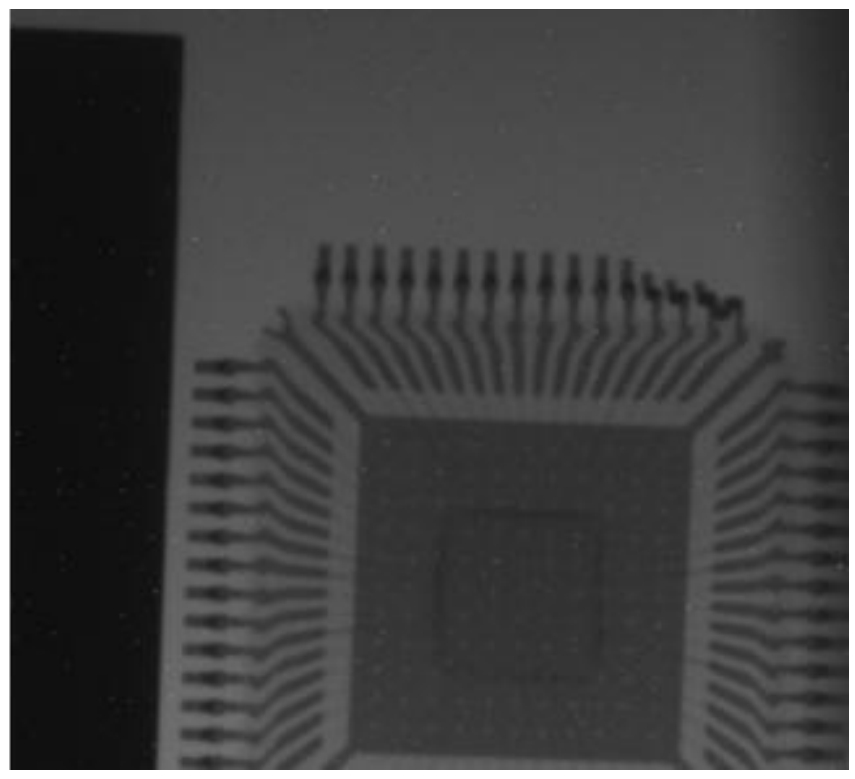

**Figure 13**. Recorded images of (left) Pb star test pattern and (right) the W-plate alongside an electronic microchip under X-ray irradiation.

### 3.3 Effect of ISO and exposure time settings

As mentioned previously, although the RPi HQ camera is not designed for low-light use, its performance can be improved through optical adjustments (back-focus alignment, low f-number) and careful optimization of ISO and exposure time. ISO acts as a gain control, increasing brightness but also amplifying noise. In contrast, longer exposures collect more photons but also increase the readout noise, which could be measured from dark frames. A trade-off between sensitivity and noise makes systematic testing essential. To investigate this, slanted-edge images of the W-target under X-ray irradiation were recorded at various ISO and exposure settings, then their corresponding histograms, contrast values were extracted.

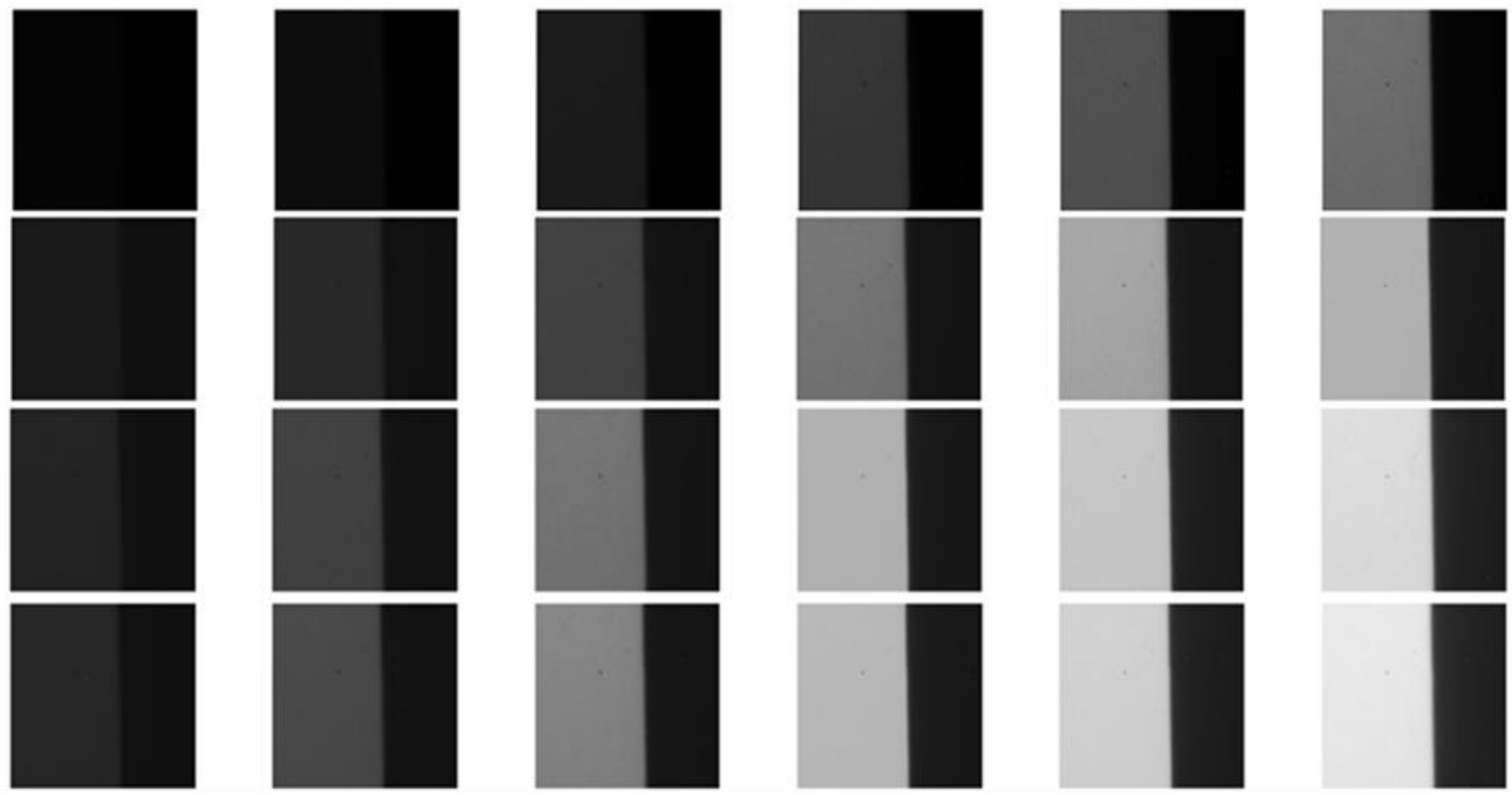

**Figure 14**. Slanted images recorded under X-ray irradiation with various ISO-exposure time settings: from top to bottom: ISO100, 200, 400 and 800, from left to right: exposure time from 100 – 2000 ms.

The images obtained at various ISO-exposure time settings are displayed on Fig. 14. As clearly shown, higher ISO and longer exposure both increase brightness, confirming that a greater dose yields a stronger scintillation signal. Contrast curves measured from fixed ROIs with various exposure times and ISO are plotted in Fig. 15. Contrast rose gradually with exposure to low ISO but reached a plateau beyond 500 ms at higher ISOs. ISO 400 at 500 ms provided approximately 73% contrast with efficient exposure and manageable noise, making it well-balanced for dose-limited or dynamic X-ray imaging. With these optimized settings, the device was further evaluated using a standard hospital X-ray machine, varying tube current and exposure time to study their effects on contrast, SNR and MTF.

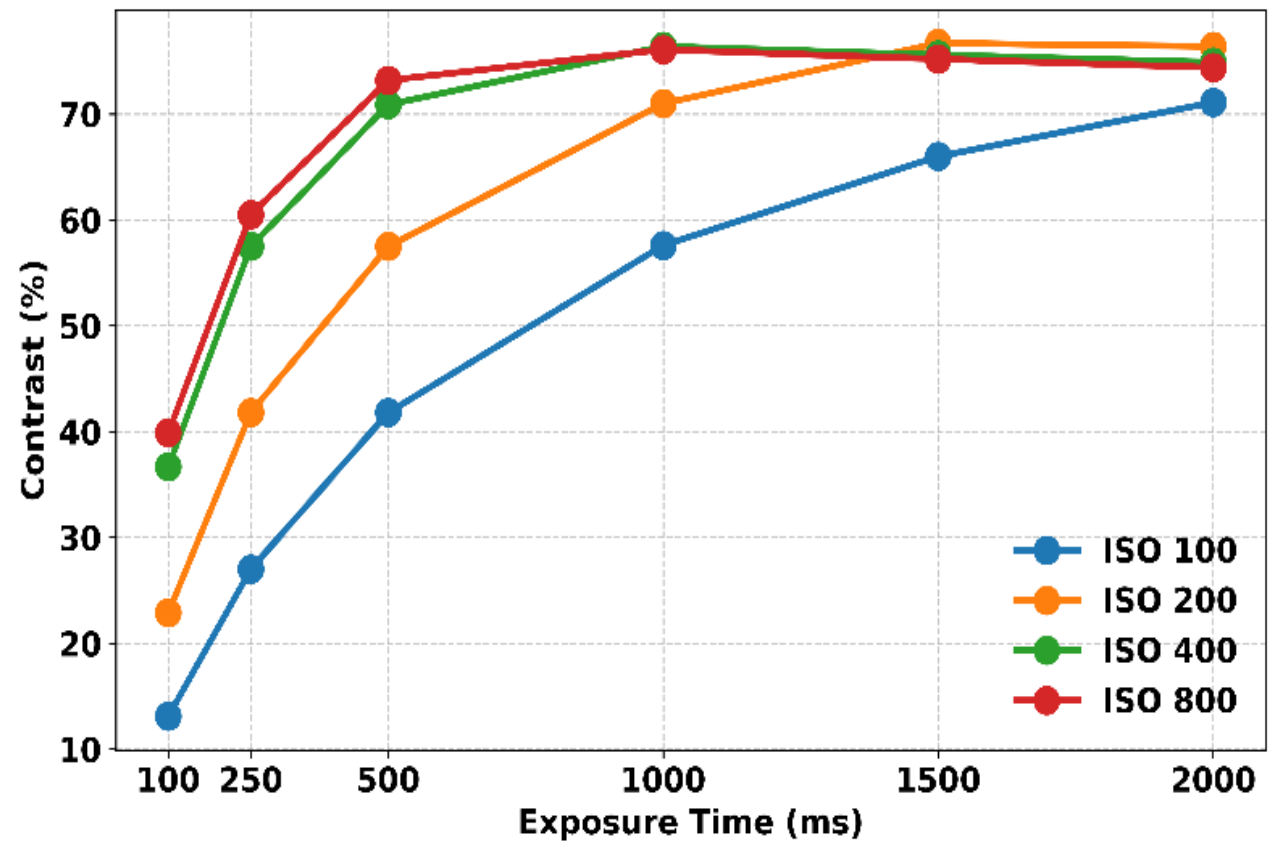

**Figure 15**. The relationship between image contrast and ISO-exposure times of recorded images under 70 kV-2 mA X-ray.

### 3.4 Modulation Transfer Function

To quantify spatial resolution, slanted-edge images of the W target under 50 kV X-ray exposure were processed to extract the ESF and LSF, as shown in Fig. 16. The analysis incorporated the camera's 1.55 μm pixel size during Fourier transformation, enabling accurate MTF curves across a wide spatial frequency range. These results establish a baseline for evaluating system performance under varying X-ray conditions. Figs. 17 and 18 illustrate the strong dependence of spatial resolution on both X-ray tube voltage (kV) and exposure level (mAs) in the indirect imaging system using a GOS scintillator.

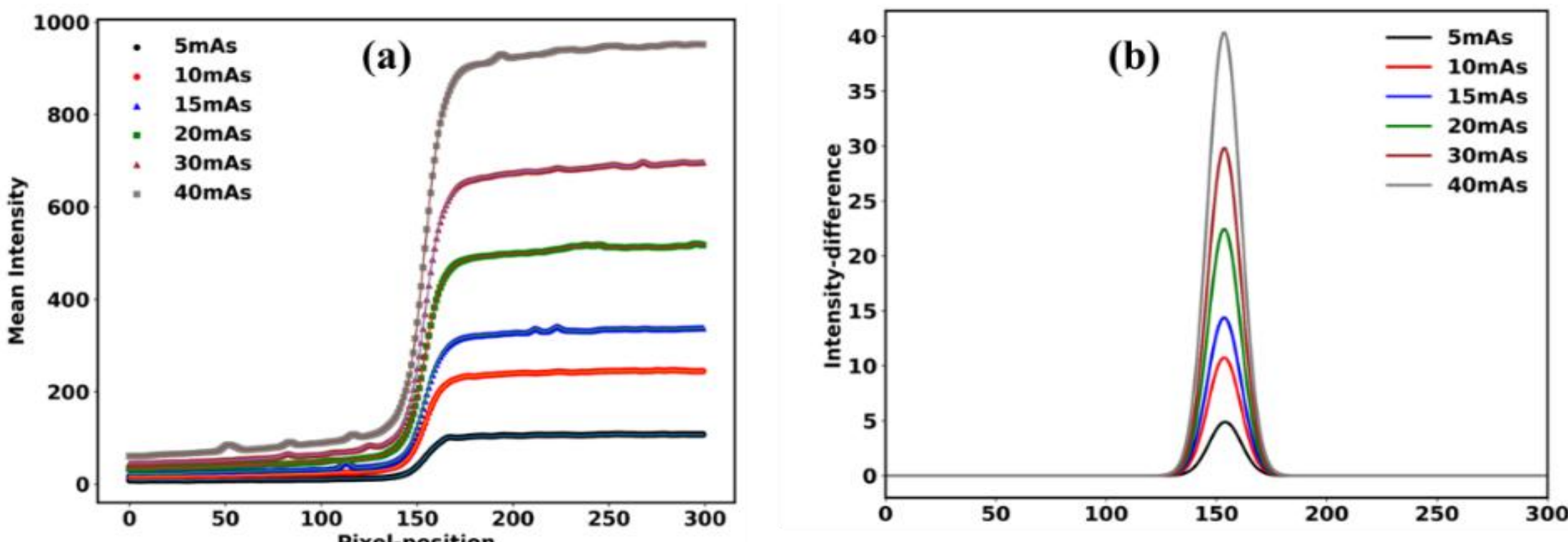

**Figure 16**. (a) The extracted Edge Spread Function and (b) corresponding Line Spread Function of the W-plate object under X-ray irradiation of 50 kV with different mAs values.

At 50 kV tube voltage, the X-ray beam has limited penetration depth and a higher absorption cross-section, causing most photons to be absorbed near the entrance surface of the scintillator [92, 93]. This shallow interaction leads to increased light scatter within the phosphor layer, degrading the MTF and reducing image sharpness. As the voltage increases (70 kV), the more energetic photons penetrate deeper into the scintillator, interacting closer to the camera sensor. This shift reduces the optical path length and internal scattering, resulting in a more confined light distribution and improved spatial resolution. This interpretation is supported by the results of MTF at 70 kV: the MTF values at high spatial frequency are slightly enhanced by increasing mAs.

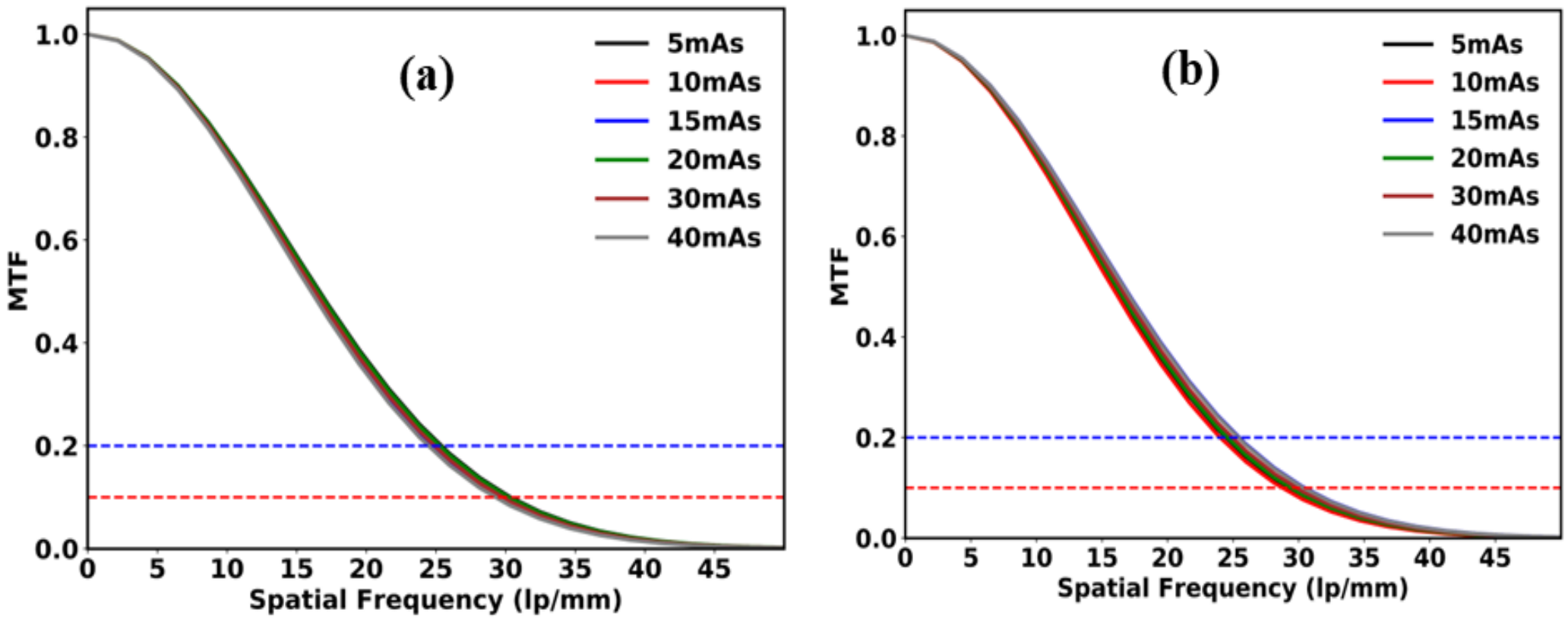

**Figure 17**. The extracted MTF curves from edge images under X-ray irradiation at (a) 50 kV and (b) 70 kV with various mAs values.

To clarify this energy-dependent relationship, we analyzed the X-ray spectra and quantified the mAs-to-exposure-dose correlation across the kilovoltage range (50 and 70 kV). Fig. 18 presents the simulated X-ray spectra for each tested voltage and the resulting exposure dose as a function of exposure level (mAs). These spectral and dosimetric analyses reveal that spatial resolution is jointly affected by tube voltage, which controls photon penetration and energy deposition, and by exposure level, which governs dose, thereby affecting signal strength. Effective imaging thus requires careful co-optimization of voltage and exposure parameters to maximize resolution within acceptable dose constraints.

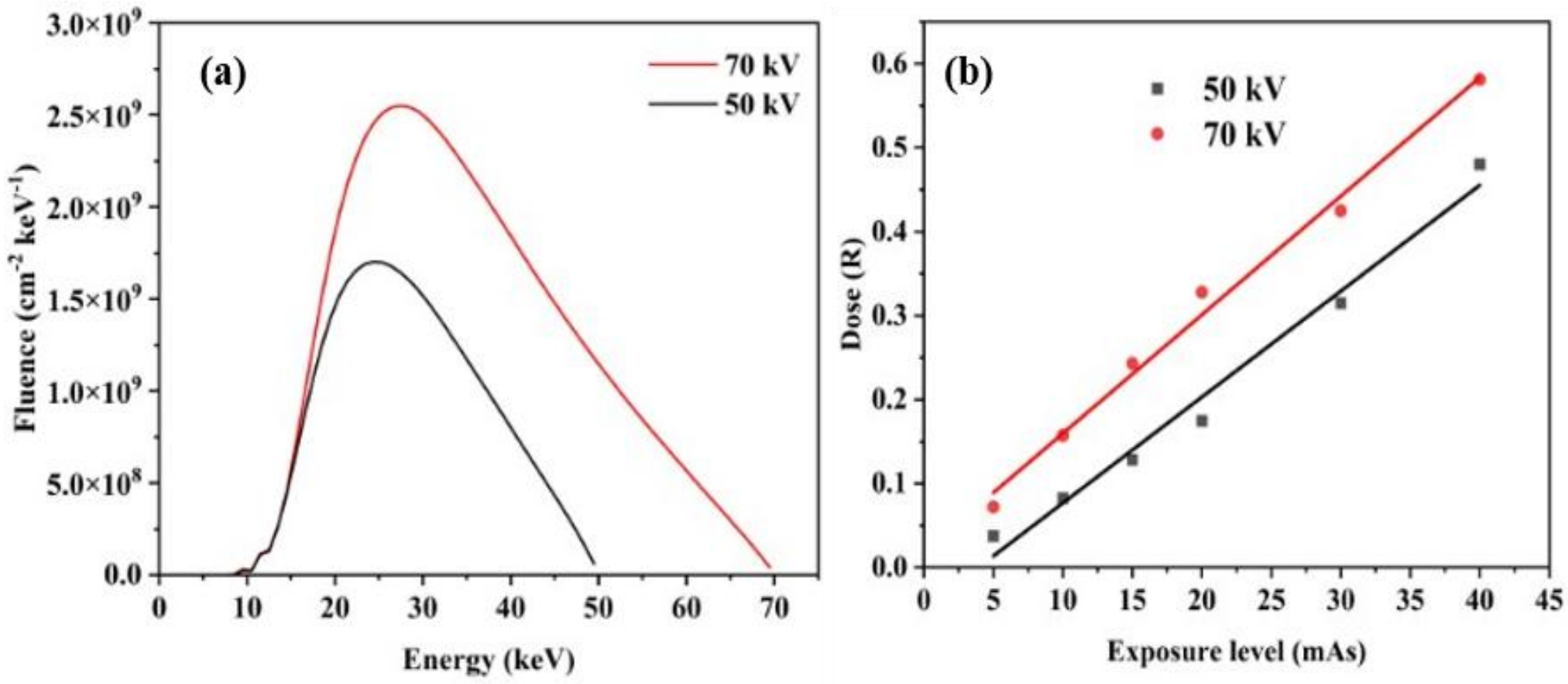

**Figure 18.** (a) Simulated energy spectrum of X-ray at various tube voltages (kV), 40 mAs by using SpekPy toolkit [94, 95] and (b), measured dose at various kV-mAs settings.

### 3.5 Image Contrast and Signal-to-Noise Ratio

Besides the MTF analysis, image contrast and signal-to-noise ratio (SNR) are strongly influenced by the X-ray tube voltage, as shown in Figs.19 and 20. The observed inverse relationship between contrast and tube voltage arises from the energy-dependent nature of X-ray interactions: low-energy photons are dominated by photoelectric absorption, enabling greater energy deposition within the medium, whereas high-energy photons interact primarily through Compton scattering, resulting in reduced energy transfer and lower light output. At lower tube voltage (50kV), The X-ray spectrum is weighted toward lower photon energies where photoelectric absorption dominates. The tungsten target acts as an effective spectral filter, strongly attenuating low-energy photons through photoelectric interactions while transmitting only a reduced fraction of higher-energy photons. Consequently, the scintillator region beneath the tungsten receives both a reduced photon count and photons with higher average energy, resulting in less energy deposited per photon, which further weakens the scintillation output. In marked contrast, the uncovered region receives the full, unattenuated photon flux, enriched in low-energy photons that interact more efficiently with the scintillator, resulting in substantially greater emissions. This combined effect, both photon number and energy transfer efficiency, yields high radiographic contrast, with the tungsten-covered region appearing distinctly dark and the uncovered region appearing bright. However, the SNR in the attenuated region remains limited by the low total photon flux, constraining the signal quality despite excellent contrast.

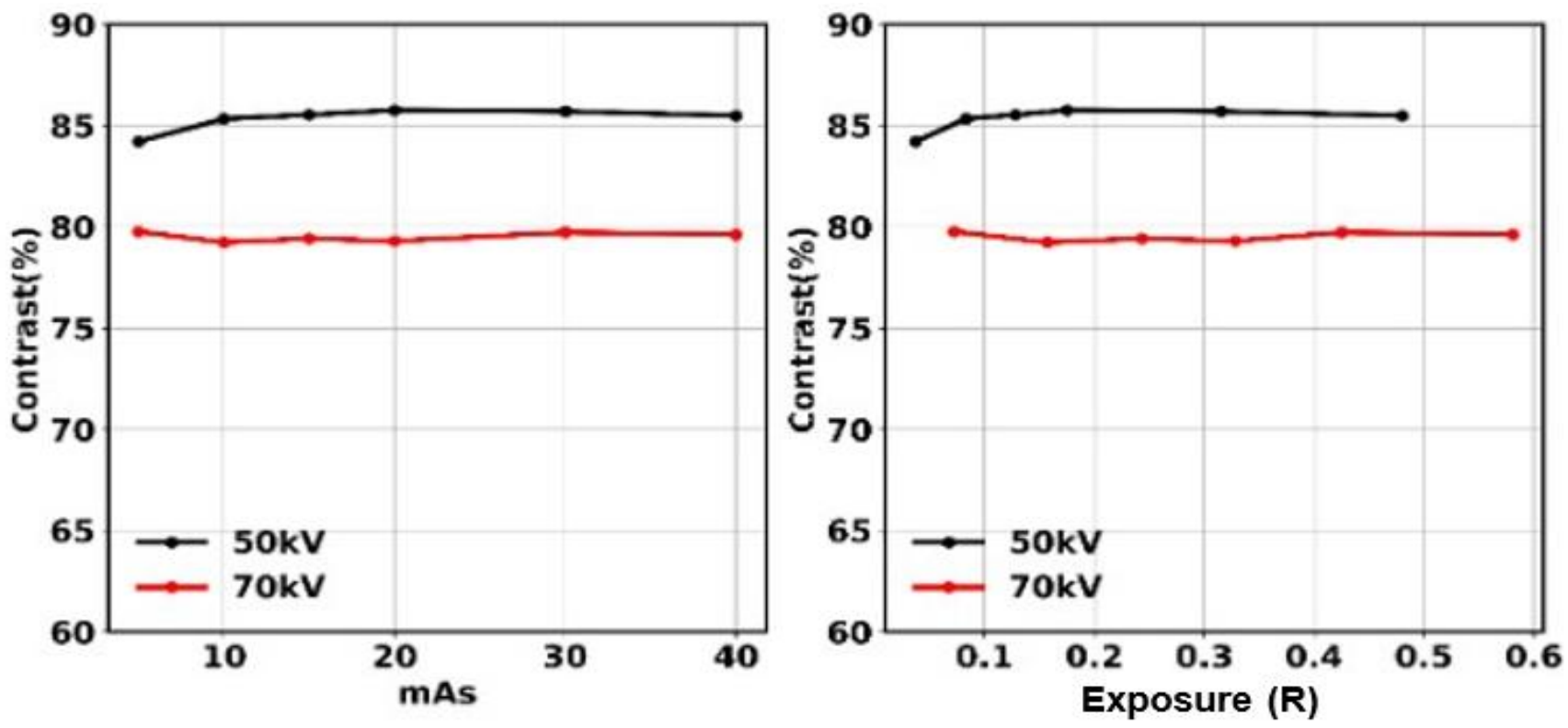

**Figure 19**. Investigation of relation between image's contrast and tube's voltage (kV) at various mAs, exposure dose.

At higher tube voltages (70 kV), the X-ray spectrum shifts toward higher photon energies, and Compton scattering becomes the dominant interaction mechanism. Compton-scattered and primary high-energy photons penetrate the tungsten target with substantially reduced attenuation due to the weak Z-dependence of Compton cross-section. However, because high-energy photons exhibit lower scintillation yield per photon, a consequence of reduced energy transfer to scintillator, the absolute light output from the attenuated region remains relatively dim despite increased photon transmission. Meanwhile, the uncovered region receives the full spectrum, thus containing a greater ratio of higher-energy photons; although these contribute less to scintillation per photon, the total incident photon flux rises significantly with tube voltage. The net result is that radiographic contrast declines, not because the attenuated region becomes brighter (it remains dim), but because the uncovered region becomes even brighter, narrowing the brightness differential. Critically, by increasing tube voltage (kV), the total photon flux will be increased, then improves the signal-to-noise ratio. Thus, SNR and contrast move in opposite directions: SNR is improved at higher tube voltages while the contrast degraded, reflecting the competing energy-dependent responses of the X-ray spectrum and scintillator luminous yield. The signal-to-noise ratio (SNR) increases with X-ray tube voltage for a fixed exposure level (mAs), as illustrated in Fig. 20.

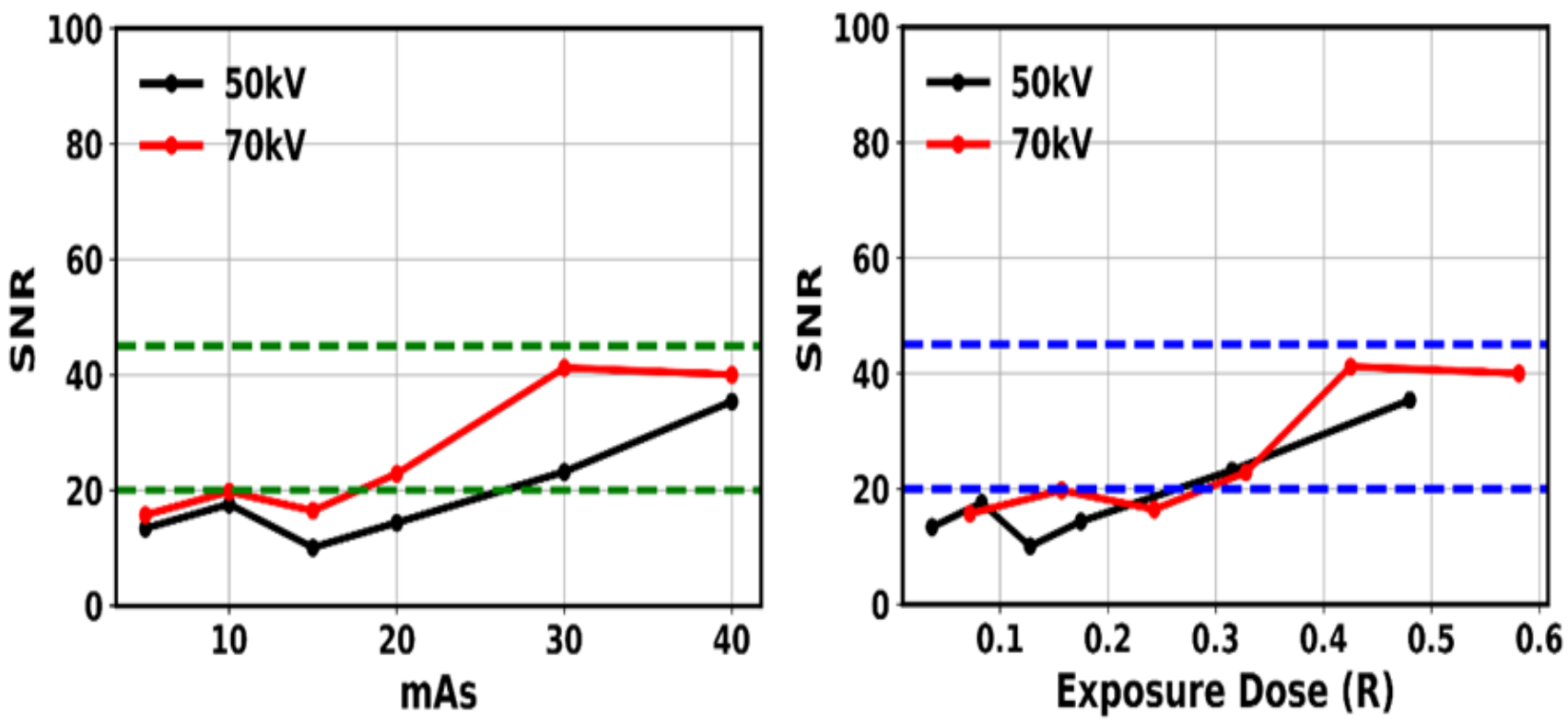

**Figure 20**. Investigation of dose efficiency in terms of Signal to Noise Ratio (SNR) versus mAs setting and measured dose at various kV.

Overall, at the same exposure level (mAs) setting, higher tube voltages provide a larger photon flux and greater beam penetration, allowing more photons to reach the scintillator and contribute to image formation. This behavior reflects the fundamental trade-off in radiographic imaging: low tube voltages produce higher contrast due to efficient photoelectric absorption of low-energy photons, while high tube voltages enhance penetration and SNR through the predominance of high-energy, Compton-scattered photons. Representative X-ray images of the tungsten target and an electronic component, demonstrating the system's imaging performance across different exposure and voltage settings is illustrated in Fig. 21. Moreover, to demonstrate the modularity of the developed device, preliminary X-ray images were acquired using alternative commercial scintillators: LYSO:Ce and GAGG:Ce. Fig. 22 presents representative X-ray images of the tungsten target recorded with each scintillator at identical acquisition settings (70 kV - 40 mAs). Clear images and visible fine structures confirm that the platform readily accommodates different scintillation screens without modification, validating the modular design approach.

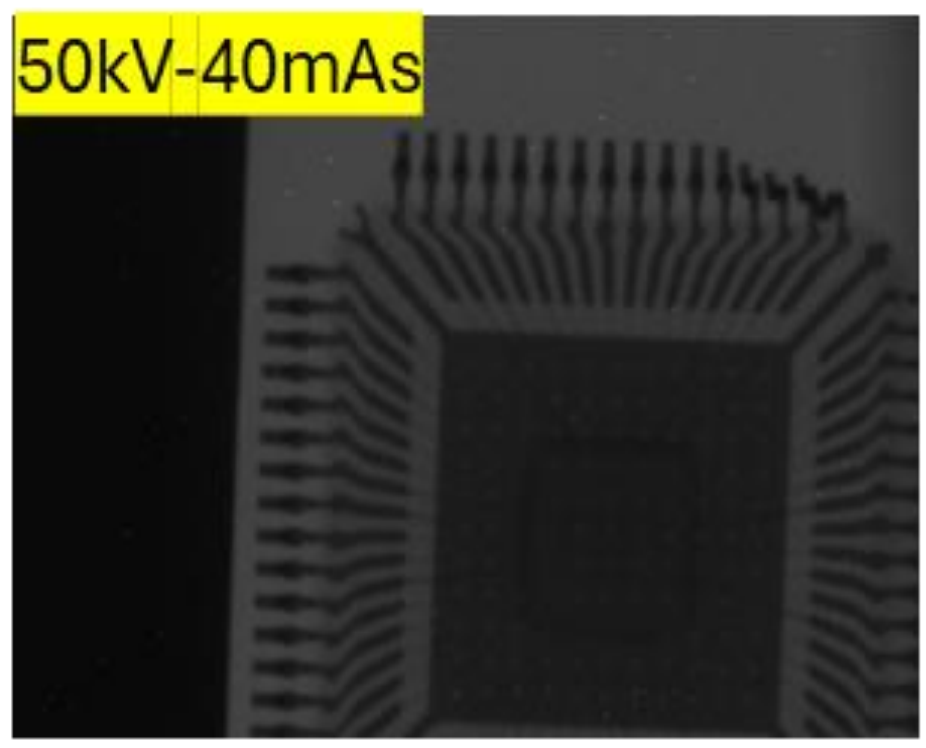

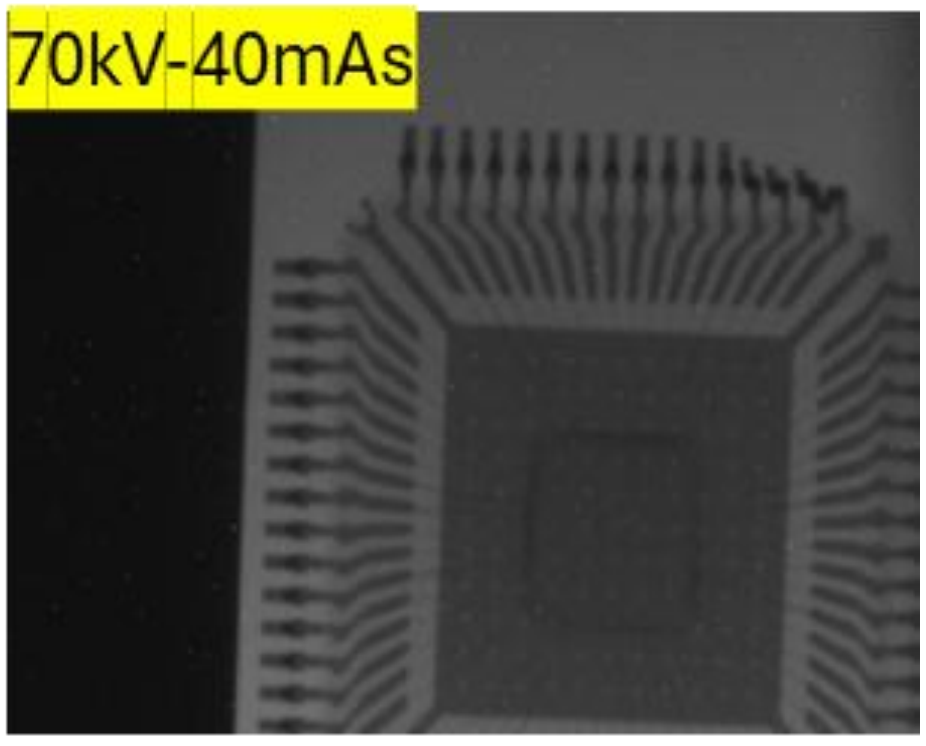

**Figure 21**. The recorded images of W target (dark area) alongside an electronic chip at X-ray of 50 and 70 kV, with a constant exposure level (tube current of 200 mA, and an exposure time of 200 ms).

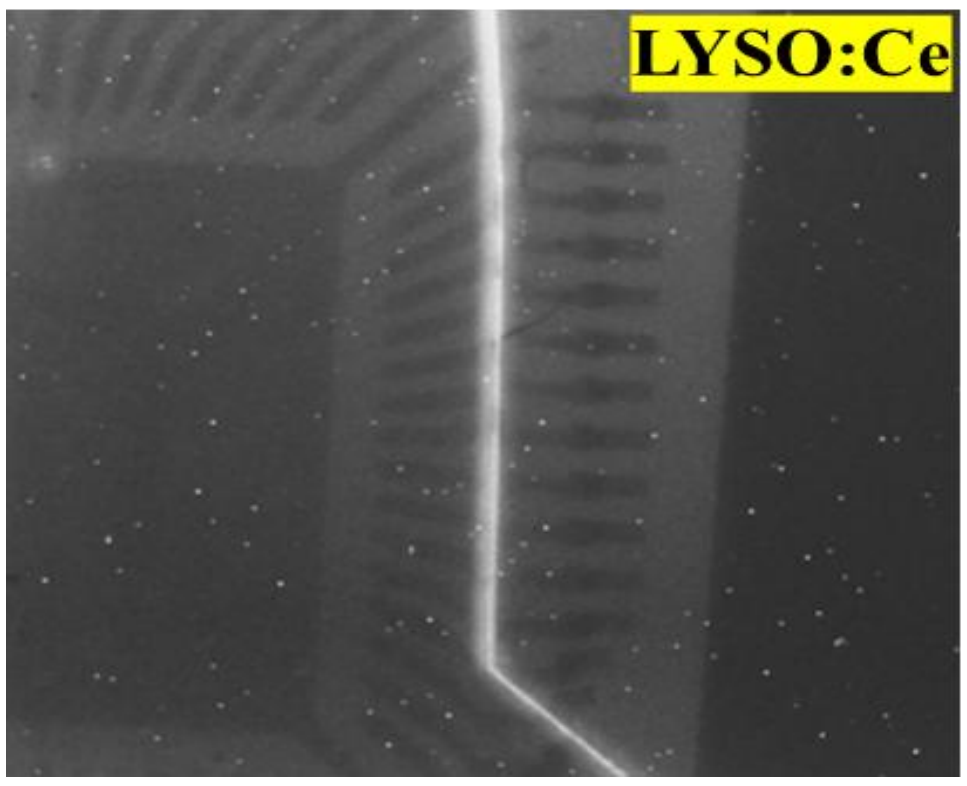

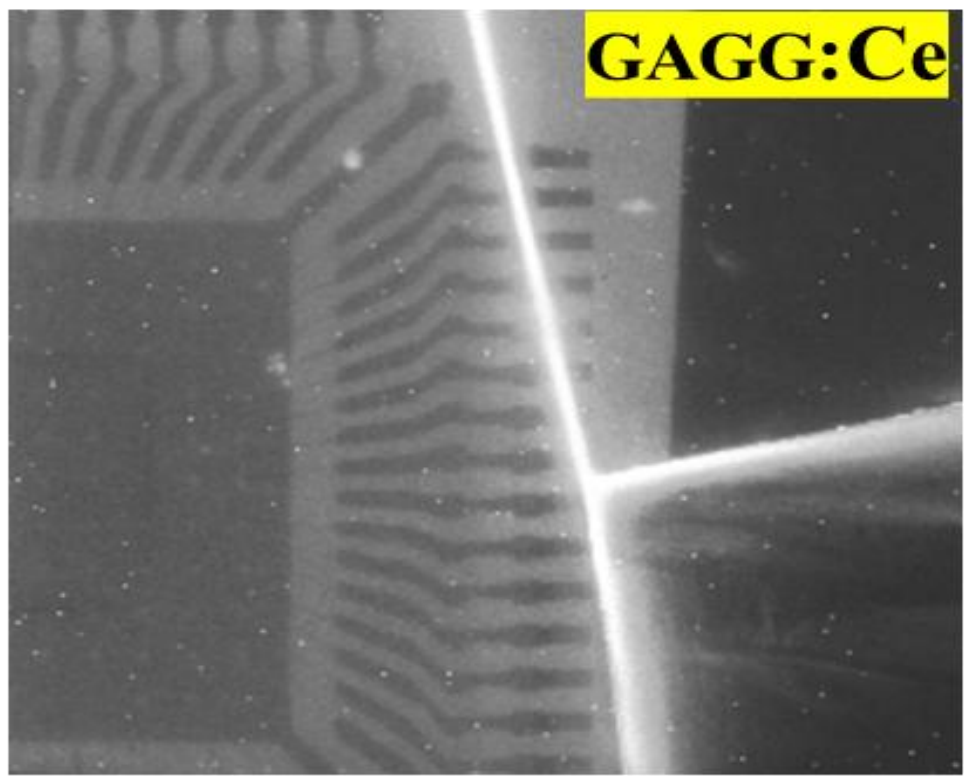

**Figure 22**. X-ray images of a tungsten (W) target positioned adjacent to an electronic chip, acquired using LYSO:Ce and GAGG:Ce scintillation screens under 70 kV - 40 mAs irradiation conditions. For enhanced visualization, the display range was adjusted to 0 - 1500 instead of the full 0-4095 range. The bright linear features observed in the images originate from cracks of the scintillation screens.

## 4 Conclusion

This study reports the development and characterization of a compact, high grade X-ray imaging system based on a Raspberry Pi computer, its high efficiency, low noise camera module, a custom optical assembly, and a GOS scintillator, with a total material cost of approximately $570. Systematic evaluation demonstrated spatial resolutions in the range of approximately 68 and 25 lp/mm ($MTF_{20}$) under ambient light and various X-ray exposure conditions. Optimization of camera parameters and exposure settings enabled effective enhancement of image contrast, signal-to-noise ratio, and spatial resolution, reaffirming fundamental radiographic principles within an accessible hardware framework.

The validated system exhibits strong technical feasibility for high-resolution digital radiography at scientific grade, substantially reduced cost. Its modular architecture allows flexible adaptation for diverse purposes, including instructional demonstrations of imaging physics, low-cost industrial non-destructive testing, and laboratory-scale research. Furthermore, owing to its indirect conversion configuration, the developed device is inherently versatile and, with an appropriate choice of scintillation screen, can be adapted for radiography applications with various excitation beams. This versatility extends its applicability to a broad range of scientific and industrial domains, such as non-destructive evaluation, materials characterization, and nuclear diagnostics. Although DQE and NNPS studies were not conducted, comprehensive characterization of resolution, contrast, and noise confirms the robustness of the imaging performance. Overall, the proposed platform provides a practical and accessible pathway toward high-resolution digital radiography and related modalities, promoting broader adoption of imaging technologies in scientific, educational, and field deployable contexts.

## Acknowledgments

These investigations were supported by the National Research Foundation of Korea (NRF), funded by the Ministry of Science and Technology, Korea (MEST) (No. RS-2018-NR031074 and RS-2024-00348317). This project is partly funded by National Research Council of Thailand (NRCT) (Grant No. N42A680125). We are also grateful to Thailand Science Research and Innovation and Nakhon Pathom Rajabhat University for supporting this research. This research was supported by Fundamental Fund (2026), Chiang Mai University. The authors would like to express their sincere appreciation to Dr. Jeongmin Park at the Korea Atomic Energy Institute for kindly providing the calibrated RadEye Remote HR system used in the initial validation experiments.